# ASPECTS OF ENTERTAINMENT DISTRIBUTION IN AN INTELLIGENT HOME ENVIRONMENT

Radu ARSINTE (*)

(*) Technical University Cluj-Napoca, Tel: +40-264-595699, Str. Baritiu 26-28, Radu.Arsinte@com.utcluj.ro

**Abstract:** The paper presents an implementation and tests of a simple home entertainment distribution architecture (server + multiple clients) implemented using two conventional cabling architectures: CATV coaxial cable and conventional Ethernet. This architecture is created taking into account the "Home gateway" concept present in most attempts to solve the problem of the "Intelligent home". A short presentation of the experimental is given with an investigation of the main performances obtained using this architecture. The experiments revealed that this simple solution makes possible to have entertainment and data services with performances close to traditional data services in a cost-effective architecture.

## I. IN-HOME TV AND DATA DISTRIBUTION

A domestic network is a part o a larger architecture. As an example, in figure 1 a typical cable network is presented. Briefly, the requirements for a domestic network are the following:
- simple and affordable;
- fully compatible with the existing analog TV equipment;
- have the possibility to be expanded to Digital technology without re-cabling or other major modifications;
- capabilities to add data services (Internet access, IP-telephony) creating a so called "triple-play" architecture
- capability to be adaptable to future digital TV technologies (DVB-C, OpenCable , IPTV)

The basic architecture for a entertainment home network is described in figure 1. The main element is the "Home Gateway" ([2]) as a bridge between the "global data community" and the "local data community" inside the house. There are four major trends that are defining consumer requirements on the home network.

*Broadband*

We have reached the stage where a majority of households understand that having a broadband connection is as much an essential part of their lives as plumbing and electricity. Not all will utilize its full capability, but broadband ubiquity opens up possibilities for business, government, schools, health care, security providers to offer services to consumers. Web 2.0 and the long tail enhance capabilities in the broadband network for social networking, communication and information sharing among users. Individuals are offered a wide choice and businesses are able to reach a large base of customers based on new distribution possibilities.

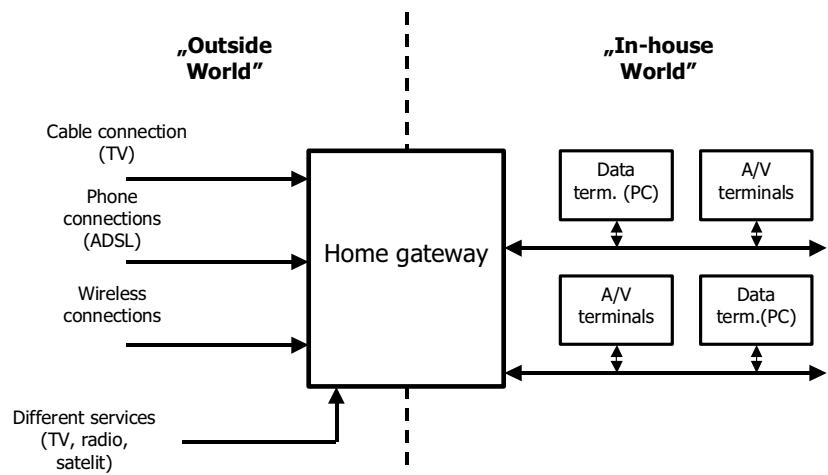

*Figure 1. In-home data and entertainment service distribution [2]*

*Complexity*

Multiple devices wish to share the broadband connection. Games consoles, PC's, telephones and IPTV set-tops want a broadband connection, and the consumer needs to be able to share that connection between all devices, simultaneously. They need a way to simply turn on the device and obtain access to network based services. Data can cross a wide variety of physical media, from radio to power line, from twisted pair to coaxial and optical fiber. The customer wants it just to work, without having to understand all the complexities in how to get it to work.

*Convergence*

The most hyped convergence service is fixed mobile convergence (FMC), where the same service will work in the mobile network and the broadband network. Initially voice services are the most important FMC services, allowing the use of your mobile phone in the home network. But there are other converged services that are equally important to the consumer, including the convergence of home networks and broadband networks, i.e. making the home network a part of the Internet. Here the key issue, in addition to simplicity, is security. And of course there is the convergence of digital media and devices, where the main issue is how to move media from one device to another, sharing it with others and while in the case of protected material, still protecting the interests of the content owners.

*Personalization*

There is a need to have personalized interaction with services and content across terminals and places. Users want the choice of services from multiple service providers combining and blending them at will to create new value. Parents want to limit their children's access to some services. To accomplish this, new enablers need to be created in the broadband network, extending from single sign-on to presence based services, to user profiling. The personal integrity of the user also needs to be maintained.

A practical wired home access alternative in many countries is the CATV network, which was initially designed to broadcast only analog TV signals from the service-provider headend to subscribers. The incorporation of fiber lines has evolved the CATV to broadband HFC plants, while enhancements in modulation and use of unused spectrum have enabled the transmission of telephony and bidirectional broadband digital signals over the legacy CATV network.

Most cable providers are offering today TV and data services (Internet +VoIP) using a rather complex technology presented in [2]. This architecture has many advantages (a reliable technology, relative low prices, a comprehensive set of services). But there are some disadvantages:
- necessity to use only "cable approved" equipment in home network architecture
- difficulty to add non-standard devices in the system

Reference [1] describes a typical view of cable network architecture and administration. This technology is known as CableHome.

CableHome specifications are intended to provide Internet Protocol (IP) - based architecture for managed home-networked services on the cable network through a DOCSIS cable modem. The CableHome architecture (figure 3) accommodates any physical and link layer home network technology that supports the transport of IP packets. This layer 1 and 2 independent architecture enables cable operators to provide services to a wide range of home networking environments. The CableHome architecture provides a defined set of requirements that support practically the whole range of services that can be delivered over cable. In order to ensure wide adoption and ease of use of this specification, CableHome technical specifications are aligned with well-known industry standards, as well as other CableLabs projects. CableHome allows the use of the existing cable operators' system infrastructure, but also provides a acceptable transition path for the deployment of CableHome over older systems. The CableHome architecture provides support for existing and future IP-based services into the home.

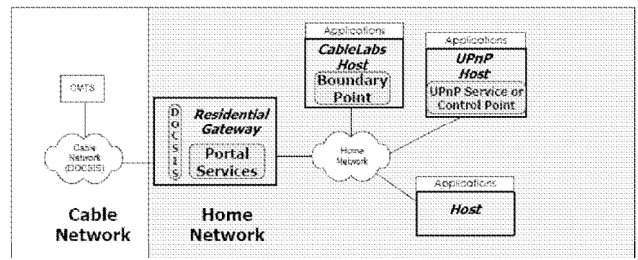

*Figure 2. Logical Reference Architecture of the CableHome system [1][3]*

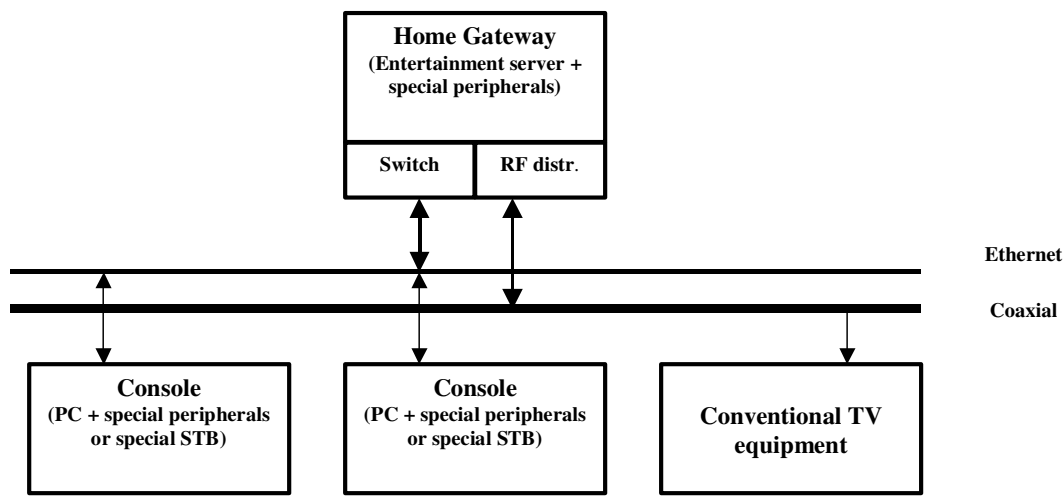

*Figure 3. Test system architecture (version 1)*

This architecture seems to have a promising future, but also being very new, it will take few years until compatible equipment could be affordable for homes. This is the main reason that leads us to investigate different options. One of such options, investigated in our research, is described in the following paragraph.

## II. HOME NETWORK ARCHITECTURE (EXPERIMENTAL SYSTEM)

Our goal was to verify if the Home network concept is useful and feasible at a complexity affordable for a normal home.

*General Architectural Concepts*

We have build two versions of the experimental system with a similar architecture.

First system (described in figure 3) uses a conventional cabling system with two sets of data paths: coaxial for analog distribution and Ethernet for data links. Normal, low cost, components were used so this architecture looks similar with a normal network, regarding data part. The A/V part is similar to normal Cable TV connections.

The second system (figure 4) is bases on the Multilet connection described in [4]. This technology makes possible to combine the proprieties of RF signal distribution and data distribution of the coaxial cable. The technology has some limitations (described also in [4]) but for an intelligent home, with limited requirements for data connections, the disadvantages are acceptable. The main advantage is the fact that the cabling is simple, or an existing cabling could be used to add data capabilities to the home. So, it seems to be a very cost-effective technology.

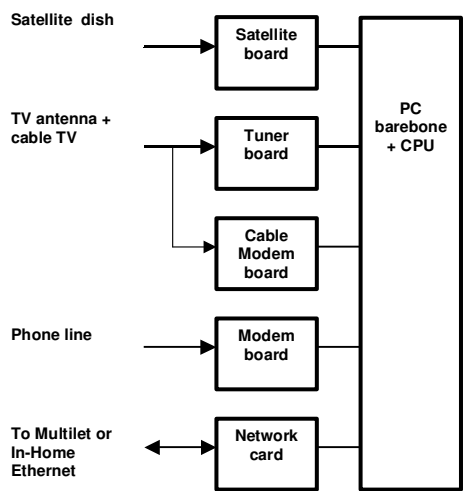

*Figure 5. Block schematic of the entertainment server*

## III. HOME ENTERTAINMENT CONTENT GENERATION

*System configuration*

To build the prototype of the entertainment server we used a conventional PC, with few multimedia capabilities. It is not necessary to have a very effective architecture since the role of the server is only to generate the content. The visualization is performed at the console.

The basic (simplified) architecture is described in figure 5.

*Software tools*

For multimedia streaming, our previous experience in this field, described in [5], was extremely useful.

The main tool used both in server and client was the well-known VLC Media Player. This program is developed in VideoLAN project. The VideoLAN project targets multimedia streaming of MPEG-1, MPEG-2, MPEG-4 and DivX files, DVDs, digital satellite channels, digital terrestrial television channels and live videos on a high-bandwidth IPv4 or IPv6 network in unicast or multicast under the main OSes.

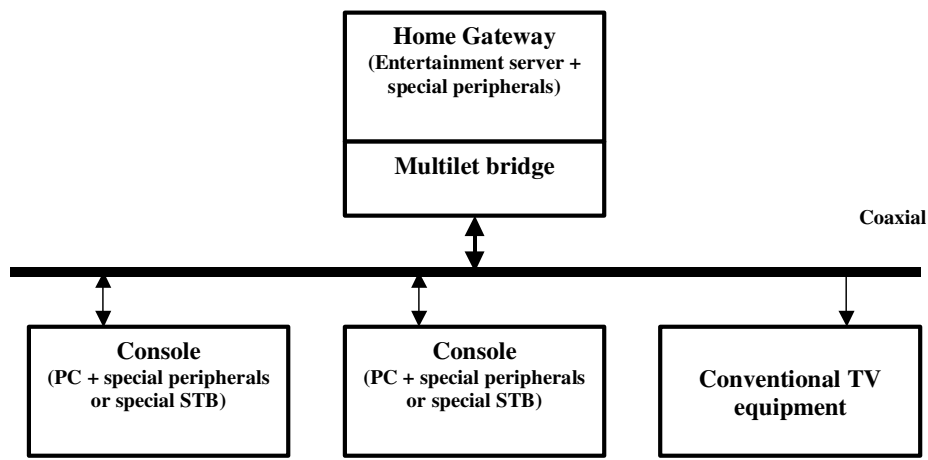

*Figure 4. Test system architecture (version 2)*

VideoLAN also features a multiple platform multimedia player, VLC, which can be used to read the stream from the network or display video read locally on the computer under all GNU/Linux implementations, all BSD versions, Windows, Mac OS X, BeOS, Solaris, QNX, Familiar Linux.

For some peripherals, with built-in support for streaming (e.g. SkyStar 2D DVB-S board) this support was also used.

For this purpose it is possible to use also Linux based software architectures, like the system described in [8].

## IV. ENTERTAINMENT CONSOLE

*Embedded system (STB –Set Top Box)*

Set-Top boxes for Ethernet are not quite common today compared with the ubiquitous STB for DVB-T or DVB-S. Consequently, the price for such STBs is relatively high. We have tested a DVB-S STB with Ethernet connectivity, having as a support a Linux operating system.

This STB called AB IP Box 200 has the following characteristics:
- Based on 300MHz PowerPC processor
- Application and Expansion. LINUX O/S enables applications and expansions.
- Max. Two Channels Recording and One Prerecorded File Playback. Simultaneous Double Descrambling with Viaccess CAM.
- Watching 1 Viaccess channel while recording another Viaccess channel with only 1 Viaccess
- CAM. Playback on prerecorded programs without Viaccess CAM.
- Variable HDD Size is Applicable for production up to 250GB or above
- Enhanced Recording Plan by timer setting/ EPG
- Enhanced Playback to skip commercials/ edit prerecorded files by PC
- USB 1.1 Supported for PC connection/ enhanced program utility/Prerecorded files uploads to PC
- Ethernet Port Supported

The performances are promising, but the firmware included in the standard delivery, makes possible only to stream data from the DVB part to Ethernet. It seems possible (according to the Internet sources) to modify the firmware and be able to accept incoming multimedia streams for visualization. We didn't explore this facility.

*Open system (PC based)*

For experiments it is more affordable to use a normal PC as a visualization console. Again, VLC media player (the client part) could perform this task. It is possible to create an entire multimedia center based on Windows technologies, as described in [6], or use Linux support like in [8]. An important issue is to optimize the data streams and the management of the incoming streams. Considerations on this matter are given in [7].

Tested performance (as presented in [4]) is excellent for one stream (one channel) and satisfactory for two streams. If we launch simultaneously more than three playbacks the limitations are visible. It is not sure that the limitations are a characteristic of the operating system's networking performance or related with VLC Media Player characteristics, but we didn't find specific works exploring this direction.

## V. CONCLUSIONS

This research make possible to evaluate the potential of a dedicated architecture in multimedia (entertainment) information creation and distribution within a normal home. The research proved that this kind of configuration is feasible and affordable, with limited features (two or three streams simultaneously). The maximum download speed for the first version (in Home Ethernet) seems to be at the limit of a normal 100Mb Ethernet. The second architecture (Ethernet over Multilet) has limited performance on data streaming, due to the limited performance of the data link delivered by the basic Multilet architecture. An important fact that must be taken also into account is the lack of the full duplex mode, due to the fundamental proprieties of Multilet. This makes impossible the simultaneous high-speed transfer of information in both directions. The main advantage of this version is the fact that the building cabling is minimal avoiding rewiring each apartment, when we need to add a new service.

The speed of the connection in version 2 is enough for most domestic multimedia and data transmission applications. For the future applications, involving high-speed Internet access, eventually IPTV, will be necessary to use version 1, or to find additional solutions based on the same concept (enhanced Multilet).

A main advantage of this architecture (using Home Gateway concepts) is the fact that it is possible to conserve for a longer period the internal structure (and investments) of the network, and apply the modifications (hardware, drivers) only in the Home Gateway, if it is necessary to adapt the network to technical progress offered by the new IT technologies.

## VI. ACKNOWLEDGEMENTS

This research is included in a CEEX research grant funded by The Romanian National Agency for Research and Technology.

This research was also possible using an evaluation system (star architecture) donated by Macab AB (Sweden).